\documentclass[12pt,preprint]{aastex}

\shorttitle{}

\shortauthors{Liu et al.}

\begin{document}

\title{Propagation and Interaction Properties of Successive Coronal Mass Ejections in Relation to a Complex Type II Radio Burst}   

\author{Ying D. Liu\altaffilmark{1,2}, Xiaowei Zhao\altaffilmark{1,2}, and Bei Zhu\altaffilmark{1,2}}

\altaffiltext{1}{State Key Laboratory of Space Weather, National Space Science Center, Chinese Academy of Sciences, Beijing 100190, China; liuxying@swl.ac.cn}

\altaffiltext{2}{University of Chinese Academy of Sciences, Beijing 100049, China}

\begin{abstract}

We examine the propagation and interaction properties of three successive coronal mass ejections (CMEs) from 2001 November 21-22, with a focus on their connection with the behaviors of the associated long-duration complex type II radio burst. In combination with coronagraph and multi-point in situ observations, the long-duration type II burst provides key features for resolving the propagation and interaction complexities of the three CMEs. The two CMEs from November 22 interacted first and then overtook the November 21 CME at a distance of about 0.85 AU from the Sun. The time scale for the shock originally driven by the last CME to propagate through the preceding two CMEs is estimated to be about 14 and 6 hr, respectively. We present a simple analytical model without any free parameters to characterize the whole Sun-to-Earth propagation of the shock, which shows a remarkable consistency with all the available data and MHD simulations even out to the distance of Ulysses (2.34 AU). The coordination of in situ measurements at the Earth and Ulysses, which were separated by about $71.4^{\circ}$ in latitude, gives important clues for the understanding of shock structure and the interpretation of in situ signatures. The results also indicate means to increase geo-effectiveness with multiple CMEs, which can be considered as another manifestation of the ``perfect storm" scenario proposed by \cite{liu14a} although the current case is not ``super" in the same sense as the 2012 July 23 event. 

\end{abstract}


\section{Introduction}

Quantifying how coronal mass ejections (CMEs), in particular fast ones, propagate from the Sun to Earth is an overarching issue in CME research and space weather forecasting. Early Doppler scintillation measurements, together with the Solwind coronagraph and Helios in situ observations, give an indication of substantial deceleration for fast CME-driven shocks (with speeds exceeding 1000 km s$^{-1}$) near the Sun \citep{woo85}. \citet{gopalswamy01a} introduce an average cessation distance of 0.76 AU (about 163 solar radii) from the Sun, after which CMEs move with a constant speed, in order to match coronagraph observations near the Sun and in situ measurements at 1 AU. The extent and cessation distance of the deceleration cannot be precisely determined given a single point of Doppler scintillation measurements in the former study and in the latter the large data gap between the Sun and 1 AU. \citet{reiner07} argue that CME deceleration can cease anywhere from about 0.3 AU to beyond 1 AU based on a statistical analysis of interplanetary type II radio bursts, but their results indeed indicate that faster CMEs decelerate more rapidly and for shorter time periods. Using a triangulation technique based on the wide-angle heliospheric imaging observations from the twin STEREO spacecraft, \citet{liu13} obtain the whole Sun-to-Earth propagation profile of fast CMEs, which typically shows three phases: an impulsive acceleration, then a rapid deceleration out to 40-80 solar radii, and finally a nearly constant speed or gradual deceleration. \citet{liu16} also find that faster CMEs tend to decelerate more rapidly and have shorter cessation distances for the deceleration, which is consistent with the results of \citet{woo85} and \citet{reiner07}. \citet{wood17} conclude that 80\% of the CMEs they studied reach final speeds before 65 solar radii by fitting the wide-angle imaging observations from a single STEREO spacecraft, confirming the results of \citet{liu13, liu16}. Now it is getting clear that a typical fast CME would finish their major deceleration well before reaching 1 AU. 

The actual situation of CME Sun-to-Earth propagation, however, may involve interactions with the highly structured solar wind including other CMEs, which complicates the issue. Interactions with other CMEs can change both the speed and propagation direction of the interacting events \citep[e.g.,][]{gopalswamy01b, farrugia04, liu12, liu14b, lugaz12, temmer12, shen12, mishra15}. Almost all studies of CME-CME interactions have focused on the interaction between two events. Interactions involving more than two CMEs can take place in the Sun-Earth space, because the typical transit time from the Sun to 1 AU is a few days, during which many eruptions could occur. Such a case has been presented by \citet{liu12}, where two CMEs from 2010 August 1, which were about 5 hr apart, merged first before they overtook a 2010 July 30 CME near the Earth. This study is enabled by STEREO's wide-angle heliospheric imaging observations, which allow to follow patterns of the interacting process continuously over essentially the entire Sun-Earth distance. \citet{liu14a} show another case involving three CMEs but with a different scenario: two closely launched events from 2012 July 23 interacted en route from the Sun to 1 AU; another CME from 2012 July 19, although non-interacting with the July 23 complex ejecta, removed some of the upstream solar wind plasma and led to a very minor deceleration of the later complex ejecta. This indicates that a non-interacting CME may also have effects on the propagation of other events by ``preconditioning" the interplanetary medium \citep{liu14a}. The importance of this preconditioning effect for CME propagation, which can last 2-5 days \citep{temmer17}, has been confirmed by \cite{temmer15} and \citet{cash15}. 

In the absence of wide-angle heliospheric imaging observations, a long-duration type II radio burst may provide a useful means to investigate the Sun-to-Earth dynamics of a CME/shock. A type II burst, typically drifting downward in frequency, is plasma radiation produced by shock-accelerated electrons near the local plasma frequency and/or its harmonics \citep[e.g.,][]{nelson85, cane87}. The frequency drift results from the decrease of the ambient plasma density when the shock moves away from the Sun. Therefore, a type II burst is usually used to derive the shock propagation distance by assuming a density model \citep[e.g.,][]{reiner07, liu08, liu09a, feng12, hu16, zhao17}. Type II emissions are also used to study CME-CME interactions, a typical signature of which is an increase in the bandwidth and intensity of the type II burst \citep{gopalswamy01b}. \citet{juan12} present an occasion of two interacting CMEs, where the type II burst appears to split into two bands with different drifting rates. Based on the study of the 2012 July 23 complex eruptions, \citet{liu17} define a complex type II burst which is composed of multiple branches that may not all be fundamental-harmonic related. In particular, they show that the low-frequency bands of the associated type II burst are consistent with the reduced plasma density ahead of the shock due to the preconditioning by an earlier CME. Despite these studies, it is still unclear how type II burst behaviors are connected with the propagation characteristics of CMEs.     

This paper attempts to address this critical question by combining coordinated coronagraph, radio and multi-point in situ observations for three successive CMEs that occurred on 2001 November 21-22. We illustrate how the characteristics of the associated type II radio burst help resolve the propagation and interaction complexities of the CMEs when wide-angle heliospheric imaging observations are not available. The paper is organized as follows. We describe the CMEs in Section 2, in situ measurements at different locations in Section 3, and the complex type II burst and the overall propagation profile in Section 4. The results are summarized and discussed in Section 5. This study provides insights into the dynamics of CME propagation and interaction in the corona and interplanetary space, the nature of radio source regions as well as space weather forecasting. It may also lead to greater use of type II radio bursts in event analysis. 

\section{CMEs near the Sun}

Figure~1 displays the coronagraph observations of the three CMEs of interest from SOHO. The first CME (CME1) occurred on 2001 November 21 and was associated with a long-duration C4.7 flare from NOAA AR 09704 (S14$^{\circ}$W19$^{\circ}$), which peaked at 14:58 UT on November 21. The second and third ones (CME2 and CME3) were launched from the Sun within 3 hr on November 22. They were associated with an M3.8 flare from AR 09698 (S25$^{\circ}$W67$^{\circ}$) peaking around 20:36 UT on November 22 and an M9.9 flare from AR 09704 (S17$^{\circ}$W36$^{\circ}$) peaking at 23:30 UT on November 22, respectively. We use a graduated cylindrical shell (GCS) method to estimate CME propagation direction and speed, which assumes a rope-like morphology for CMEs with two ends anchored at the Sun \citep{thernisien06}. The coronagraph images can be well reproduced by this forward modeling technique, but since there is only a single vantage point large uncertainties in the resulting parameters may arise. CME1 has two fronts. We choose to model the preceding one in order to have an estimate of the leading-edge speed. 

Figure~2 is an idealized 3D view of the three CMEs relative to the locations of the Earth and Ulysses. The average propagation longitude of CME1 is about W25$^{\circ}$, consistent with its solar source longitude (W19$^{\circ}$). CME2 and CME3 have an average longitudinal direction of about W50$^{\circ}$ and W12$^{\circ}$, respectively, so they propagate more toward the Earth compared with their solar source longitudes (W67$^{\circ}$ and W36$^{\circ}$). All the three CMEs appear to have a relatively low propagation latitude (about S11$^{\circ}$, S18$^{\circ}$ and N08$^{\circ}$ respectively) but a large flux-rope tilt angle (about 66$^{\circ}$, 101$^{\circ}$ and 46$^{\circ}$ counterclockwise from the RT plane to the north). The Earth probably encountered the flank of CME2 given its propagation longitude. A linear fit of the GCS leading-front distances gives an average speed of about 800 km s$^{-1}$ for CME1, about 1700 km s$^{-1}$ for CME2, and about 2200 km s$^{-1}$ for CME3 near the Sun. Based on these estimates and the CME launch times, we expect that CME2 and CME3 would begin to interact near the Sun. CME3 seemed to indeed propagate into the material of CME2 as can be seen from the coronagraph observations (Figure~1). The interaction with CME1 would take place far away from the Sun, as CME1 occurred more than a day earlier.      
 
\section{In Situ Measurements at Different Locations}

Figure~3 shows the in situ measurements near the Earth. Two intervals with interplanetary CME (ICME) signatures can be identified from the data. The identification of two intervals agrees with the study of \citet{rodriguez08}, except that our intervals are slightly larger than theirs. The first interval (ICME1) is probably the corresponding structure of CME1 at 1 AU, and the second one is likely a complex ejecta \citep{burlaga02} formed from the merging of CME2 and CME3. The first half of the second interval exhibits signatures of a magnetic cloud \citep[MC;][]{burlaga81}. The association with the solar sources is consistent with the arrival times of the CMEs at the Earth estimated from their launch times and speeds near the Sun. Below we will show how type II burst observations provide further support for this association. However, it is not possible to separate the interplanetary manifestations of CME2 and CME3 within the complex ejecta (second interval). The duration of the complex ejecta is about 27 hr, corresponding to a radial width of about 0.46 AU. ICME1, with a time period of about 4 hr which corresponds to a radial size of about 0.08 AU, appeared in the sheath of the complex ejecta. This is often thought to be part of a sheath region (i.e., not an ICME), but the presence of an ICME in a sheath is not rare \citep[e.g., ][]{liu12, liu14a, mostl12}, and the magnetic field inside ICME1, although turbulent, gives an indication of rotation. Further evidence for the identification as an ICME will be provided below with coordinated Ulysses measurements at 2.34 AU.

A strong forward shock with a speed of about 1050 km s$^{-1}$ passed Wind at 05:53 UT on November 24. As will be seen later, this is likely the shock which survived from the interaction between CME2 and CME3 (i.e., the merger of the shocks from CME2 and CME3). A careful look at the data reveals a small shock slightly ahead of the strong one (04:55 UT on November 24), which was probably generated by CME1. These signatures seem to suggest that at 1 AU the shock originally driven by CME3, after having passed through CME2 and CME1, was just about to exit from the sheath of CME1 and to merge with its shock. The density, speed, temperature and magnetic field strength are greatly enhanced inside ICME1, with the peak magnetic field of about 69 nT. A possible explanation for this enhancement is the compression by the second shock when it was propagating through ICME1, plus the compression by the complex ejecta from behind. A similar scenario is observed in the 2012 July 23 event \citep{liu14a}, although the magnetic field in the current case is not as high. The maximum southward magnetic field is about 50.5 nT. Although only for a brief time period, the southward field caused an intense geomagnetic storm with the minimum $D_{\rm st}$ index of $-221$ nT. The measurements also show a reverse wave around 00:52 UT on November 26 at the wake of the complex ejecta, with a weak increase in the speed and slight decreases in the density, temperature and magnetic field. 

Ulysses observed a strong shock at 15:07 UT on November 26 at a distance of 2.34 AU, latitude of $73^{\circ}$ and longitude of $47^{\circ}$ in the heliographic inertial system. It was about $57.9^{\circ}$ west and $71.4^{\circ}$ north of the Earth. For comparison, the three CMEs have an average propagation longitude of about W25$^{\circ}$, W50$^{\circ}$ and W12$^{\circ}$, respectively, and a relatively low propagation latitude with respect to the Earth (see Figure~2). The in situ measurements at Ulysses are shown in Figure~4. The overall plasma and magnetic field profiles are similar to those observed near the Earth: a complex ejecta with its first half interval looking like an MC, an ICME appearing in the sheath of the complex ejecta, enhanced alpha-to-proton density ratio inside the complex ejecta, enhanced density in the middle of the complex ejecta, similar magnetic field rotation in the MC, and a forward shock and a reverse wave around the intervals. These similarities suggest that Ulysses observed the same events despite the large latitudinal separation. The complex ejecta, again, was likely formed from the merging of CME2 and CME3. The magnetic structure of ICME1 is now much clearer than that at the Earth, resembling a small MC except that the temperature is not depressed and the field components are still turbulent to some extent. This gives further evidence that the first interval at the Earth is not a normal sheath region, but a shocked ICME.

The shock at Ulysses has a speed of about 900 km s$^{-1}$, comparable to the speed at the Earth (1050 km s$^{-1}$), so the major deceleration of the shock must have occurred inside 1 AU, and further out the shock moved with a roughly constant speed. This should be the shock that eventually survived from the interactions between the three CMEs, as there is not another shock ahead of it as at 1 AU. The reverse wave at 19:41 UT on November 29 seemed produced by the overpressure inside the complex ejecta associated with the strong magnetic field. At Ulysses the radial width is about 0.11 AU for ICME1 and about 0.87 AU for the complex ejecta, larger than the counterpart at the Earth. They might be still expanding during transit from the Earth to Ulysses, or Ulysses crossed different parts of these ejecta. Similar conclusions have been reached by \citet{rodriguez08} and \citet{reisenfeld03} on the identification of the ICMEs/shocks, although they do not recognize some of the features discussed in the present paper.       

To connect the in situ measurements at the Earth and Ulysses, we propagate the solar wind data outward from the Earth using an MHD model developed by \citet{wang00}. The model assumes spherical symmetry, since solar wind measurements are only a 1D cut. Hourly averages of the data from Wind, which smooth out the small shock, are used as input to the model. The results are displayed in Figure~5. The large preceding shock persists out to Ulysses. The predicted arrival time of the shock at Ulysses is only about 1.2 hr earlier than observed. This time difference is much smaller than the propagation time of about 57.2 hr from the Earth to Ulysses. The good alignment provides further evidence that Ulysses and the near-Earth spacecraft observed the same events. It also confirms that the shock at Ulysses is the final shock which survived from the interactions between the three CMEs. Note that the ambient solar wind speed from the model is much lower than observed at Ulysses. This is reasonable as Ulysses was near the north pole ($73^{\circ}$).

\section{Complex Type II Burst and Overall Propagation Profile}

Figure~6 shows the radio dynamic spectrum associated with the CMEs. A narrowband type II burst from CME2 first appeared. Then from about the beginning of November 23 a long-duration type II burst occurred with a large bandwidth. \citet{gopalswamy03} suggest that this is a broadband enhancement produced by the interaction between CME2 and CME3. Since the speeds of the two CMEs (1700 and 2200 km s$^{-1}$ respectively) are comparable, their interaction and thus the type II burst enhancement would continue for a long time. A further, sudden enhancement in the bandwidth and intensity of the type II burst was observed at 00:07 UT on November 24. A first impression about this sudden enhancement is that a shock may have arrived at the spacecraft, but the in situ measurements at Wind/ACE do not show any shock around that time (see Figure~3). We propose that the enhancement in the type II burst resulted from the overtaking of CME1 by the shock which had emerged from the interaction of the later two CMEs. The distances traveled by the CMEs support this interpretation: at 00:07 UT on November 24 CME1 would have reached $0.9\pm0.1$ AU from the Sun, which is similar to the distance of about 0.85 AU traveled by the shock (see Figure~7). The distance of CME1 is estimated from the range of its speed in the Sun-Earth space, i.e., about 800 km s$^{-1}$ near the Sun and about 600 km s$^{-1}$ at 1 AU (the speed of the first shock in Figure~3). Another intensification in the type II burst occurred around 05:53 UT on November 24, which is indeed the shock arrival at the spacecraft (see Figure~3). The intensity of $>$5 MeV particles plotted in Figure~6 exhibits two peaks, both of which are coincident with the enhancements in the type II burst. The second peak is due to the trapping of particles around the shock known as energetic storm particles, while the first one must indicate a strengthening of the shock.  

We use a kinematic model to characterize the shock propagation and simulate the frequency drift of the Sun-to-Earth type II burst. The shock is assumed to start with an initial speed $v_0$ at a time $t_0$ and a distance $r_0$ from the Sun, move with a constant deceleration $a$ for a time period $t_a$ before reaching 1 AU, and thereafter travel with a constant speed $v_s$ as measured at 1 AU. The initial speed $v_0$ (2200 km s$^{-1}$) and distance $r_0$ are derived from GCS modeling which removes projection effects. Here we assume that the shock nose distance is not much different from that of the CME leading edge, which is usually true in the coronagraph observations. The shock speed $v_s$ (1050 km s$^{-1}$) and transit time $t_s$ are known from in situ measurements at 1 AU. The whole propagation profile can then be uniquely set without any free parameters. For simplicity we take $r_0\sim 0$ (which is actually a few solar radii) and $t_0$ to be the mid-time between the flare start (22:32 UT on November 22) and maximum (23:30 UT on November 22). With these known parameters we obtain the time for deceleration $t_a=2(d_s-v_st_s)/(v_0-v_s)$ where $d_s$ is the distance of Wind/ACE from the Sun, the deceleration $a=(v_s-v_0)/t_a$, and the cessation distance of deceleration $r_a=(v_0+v_s)t_a/2$. This formulation gives a simple analytical description of the whole propagation profile. For the current case these expressions yield $a=-20.8$ m s$^{-2}$, $t_a=15.4$ hr and $r_a=0.6$ AU.   

The distances resulting from the analytical model are then converted to frequencies using the density model of \citet{leblanc98}. The Leblanc density model is appropriate for a distance range from about 1.8 solar radii to 1 AU, which is almost the same as that covered by the frequencies of the dynamic spectrum. In order to match the band of the type II burst, we adjust the density scale $n_0$ in the density model, the nominal density at 1 AU. A value of $n_0=16$ cm$^{-3}$ and the assumption of type II emission at the second harmonic of the local plasma frequency give a frequency-time profile that is well consistent with the observed type II band over essentially the whole Sun-to-Earth time range (see Figure~6)\footnote{Here the purpose is to show that the analytical results can be consistent with the radio data, and this consistency is actually independent of the emission mode assumed.}. Note that the density model describes the average radial variation of the density of the medium where the shock was propagating, so the scale factor $n_0$ is not necessarily the observed density at 1 AU. This analytical approach is different from previous studies which rely on a fit of the frequency drift of a type II burst \citep[e.g.,][]{reiner07, liu08, zhao17}. It may set up a new paradigm in characterizing CME/shock propagation with long-duration type II bursts.      

The resulting time for deceleration (15.4 hr) corresponds to the time of 14:25 UT on November 23, after which the shock moved with a constant speed. Around 14:25 UT on November 23 the type II burst indeed showed a slight but discernible change in its appearance. Presumably, this is the time when the shock emerged from the interaction between CME2 and CME3. Otherwise the shock would continue to slow down after that time. While the ambient solar wind could contribute to the deceleration, the major slowdown of the shock would result from its interaction with CME2. The enhanced radio emission during the CME-CME interaction indeed implies an increased density within CME2 compared with the ambient medium. Based on this rationale, we suggest that it took about 14 hr for the shock to propagate through CME2 if they began to interact around the beginning of November 23 (see Figure~1). The merging of CME2 and CME3 or the equalization of their leading-edge speeds might take about the same time scale. The merged CMEs would thereafter have a significantly reduced deceleration in the solar wind due to the increased mass and inertia from the merging. This is likely the reason that the shock had a roughly constant speed after 14:25 UT on November 23 even when it was propagating inside CME1. The propagation of the shock through CME1 only took about 5.8 hr, i.e., the time interval between the second and third vertical dotted lines in Figure~6. This inference is obtained by combining Figure~3, which shows that the shock was nearly exiting from the sheath of CME1, and Figure~6, which indicates that the shock began to overtake CME1 around 00:07 UT on November 24. The time interval of 5.8 hr corresponds to a radial width of about 0.06 AU for CME1, i.e., $(v_s-v_s^{\prime})\times5.8$ hr, where $v_s^{\prime}$ is the speed of the shock driven by CME1 at 1 AU (about 600 km s$^{-1}$). The value of 0.06 AU is comparable to the radial size of ICME1 at 1 AU (0.08 AU).                 

We extend the propagation profile of the shock out to the distance of Ulysses, as shown in Figure~7. Also plotted in the figure are the locations determined from GCS modeling of LASCO observations, the distances from the type II frequencies, the in situ measurements of the shock arrival at the Earth and Ulysses, and MHD model output in between. The analytical profile agrees with all the data. Note that the height-time profile is obtained only from the initial speed near the Sun ($v_0$) and the shock parameters at 1 AU ($v_s$ and $t_s$). Overall, the shock is tracked very well by the analytical profile, even to the distance of Ulysses (2.34 AU). This good agreement verifies the simple kinematic model for the shock propagation, in particular the assumption that the major deceleration is finished well before reaching 1 AU. The cessation distance of deceleration (0.6 AU) is smaller than that obtained by \citet{reiner07} for the same event (0.77 AU), but larger than those by \citet{liu13} for some fast CMEs (0.2-0.4 AU). In the current case, the deceleration is mainly due to the propagation of the shock inside CME2; the ambient solar wind slowed down CME2 and thus contributed to the deceleration of the shock indirectly. This is different from the cases of \citet{liu13}. The interaction between CME2 and CME3 lasted for a relatively long time because their speeds are comparable. As a result, the shock did not show an abrupt slowdown when it was propagating inside CME2, but rather a continuous deceleration. The long interaction time is also likely responsible for the large cessation distance of deceleration compared with those in \citet{liu13}. 

\section{Conclusions and Discussion}

We have investigated the propagation and interaction properties of three successive CMEs from 2001 November 21-22, using coordinated coronagraph, radio and multi-point in situ observations. A particular focus is to address how the behaviors of a long-duration complex type II radio burst are connected with the propagation and interaction characteristics of the CMEs. In addition to a consistent story on the current cases, we obtain key findings on CME coronal and interplanetary dynamics, shock structure and the nature of radio source regions, which may apply to other events. The work also presents a detailed investigation of interactions involving more than two CMEs, which has been lacking in previous studies. The results are summarized and discussed as follows. 

1. The long-duration type II burst provides key features that help resolve the propagation and interaction complexities of the three successive CMEs, in combination with coronagraph and in situ observations. The two CMEs from 2001 November 22 (CME2 and CME3) interacted first, producing a broadband enhancement in the type II burst. After having passed through CME2 the shock originally driven by CME3 then overtook the 2001 November 21 event (CME1) at a distance of about 0.85 AU from the Sun, as can be seen from another enhancement in the bandwidth and intensity of the type II burst around 00:07 UT on November 24. This scenario is similar to the case studied by \citet{liu12} which also shows interactions among three CMEs. Even without wide-angle heliospheric imaging observations as in \citet{liu12}, we are able to determine the time scale that the shock from CME3 spent inside CME2 and CME1 using the type II features, which is about 14 and 6 hr, respectively. The shock did not slow down abruptly when it was interacting with CME2, but exhibited a continuous deceleration as the speeds of CME2 and CME3 are comparable. After exiting from CME2 the shock had a roughly constant speed even during its propagation inside CME1, which can be attributed to the increased mass and inertia from the merging of CME2 and CME3. How the shock merged with each of the shocks from the two preceding CMEs, which is expected to occur instantaneously, cannot be determined from the data.

2. The work presents a simple analytical approach that may set up a new paradigm in characterizing the Sun-to-Earth propagation of fast CME-driven shocks. Although the model has been employed in previous studies \citep[e.g.,][]{gopalswamy01a, reiner07, liu08}, we demonstrate how its analytical practice gives the simplicity and ease of use without any free parameters. A simple analytical description of the whole propagation profile is achieved with the known initial speed near the Sun and shock arrival time and speed at 1 AU based on the assumption that the major deceleration is finished before reaching 1 AU. The initial speed of the CME/shock can be obtained from GCS modeling of coronagraph images. In the current case, coordinated data from LASCO, the long-duration type II burst and in situ measurements at the Earth and Ulysses as well as MHD model output between the Earth and Ulysses are all consistent with the analytical height-time profile of the shock. The fact that the shock completed its major deceleration inside 1 AU has important implications for CME research and space weather forecasting, as discussed in \citet{liu13, liu16}. The deceleration in the present case, however, mainly resulted from the interaction between the shock and CME2, which gave rise to a relatively large cessation distance of deceleration (0.6 AU) compared with those in \citet{liu13} and \citet{wood17}.  

The kinematic model can also be used to describe the propagation of the CME driver, as has been demonstrated in previous studies \citep[e.g.,][]{gopalswamy01a, reiner07, liu08, zhao17}. The shock is not a freely propagating shock but gains energy continuously from the CME, despite a concern on the standoff distance between the shock and driver. A preliminary analysis shows that the analytical approach also works well for the 2006 December 13 event \citep[see a comprehensive study of the CME by][]{liu08}, which was relatively isolated from other CMEs. In a separate work we will evaluate the effectiveness of the analytical model by examining a list of CMEs, which we require to be associated with long-duration type II bursts and measured in situ at both the Earth and Ulysses. The statistical results would certainly depend on the number of the samples that we can find.

3. The coordinated measurements at different locations give important clues for the understanding of shock structure and the interpretation of in situ signatures. The measurements at the Earth and Ulysses, which were separated by $71.4^{\circ}$ in latitude, indicate a large latitudinal extent of the ejecta and shock. Note that the GCS modeling gives a big flux-rope tilt angle for all the three CMEs. This could be the reason for the large latitudinal extent of the ejecta and shock. The in situ measurements at different locations also provide evidence for a shocked ICME in a sheath, which is often misinterpreted as a normal part of the sheath region. Since the MHD model used here assumes spherical symmetry, the successful model-data comparison implies that the shock surface on a global scale may be nearly spherical, with the center of the sphere at the Sun. It should be emphasized that, when the shock front is not far away from the Sun, the center of the shock sphere is not at the Sun \citep{liu17}. A similar conclusion on the shock structure has been obtained for the event from 2006 December \citep{liu08}, when the Earth and Ulysses were separated by 74$^{\circ}$ in latitude and 117$^{\circ}$ in longitude. The cessation distance of deceleration, however, is much smaller (only 0.36 AU) than that of the present case, although the initial speed is almost the same.  

\citet{liu17} indicate that the fast ambient solar wind near the pole may have helped to maintain a spherical shock structure on a global scale by modifying the speed of the shock flank. The shock nose, which was at a low latitude, would move faster than the flank in a uniform ambient medium. The flank, however, was experiencing a faster and less dense wind at the high latitude, so it would have a reduced deceleration compared with the nose. This is why the flank could keep the same pace with the nose at some distances from the Sun. However, it is unclear within what distance range the shock was a roughly spherical structure centered at the Sun, and it is not easy to get a clear picture of the shock structure with only two vantage points of in situ measurements. On a local scale the shock may have a distorted structure, e.g., a dimple in the shock front due to its interaction with the preexisting heliospheric plasma sheet as can be seen from both observations \citep[e.g.,][]{liu09b, liu11} and simulations \citep[e.g.,][]{riley03, manchester04}.

4. The results also indicate means to increase geo-effectiveness with multiple CMEs. To enhance geo-effectiveness requires simultaneous magnification of the southward component of the interplanetary magnetic field $B_s$, the solar wind speed $v$ and the density $n$, in order of importance. These parameters produce the dawn-dusk electric field ($vB_s$) governing the entry of solar wind energy into the magnetosphere and the solar wind momentum flux ($nv^2$) controlling the compression extent of the magnetosphere. The 2012 July 23 superstorm shows ``preconditioning" of the upstream medium by an earlier, non-interacting CME plus interaction between two later, closely launched CMEs \citep{liu14a}; all the three parameters of the later complex ejecta were thus enhanced simultaneously. In the present case, it is not clear whether CME1 played a role similar to the ``preconditioning". The average frequencies of the type II band seem to indicate a relatively large density upstream of the shock segment(s) that produced the radio emissions. This is probably because the shock from CME3 was largely propagating inside the preceding CMEs that had an enhanced density compared with the ambient medium. The increased mass and inertia resulting from the merging of CME2 and CME3, however, can indeed help reduce the deceleration of the merged ejecta, and hence the deceleration of the preceding shock. Interactions between the three CMEs, in particular the passage of the shock through CME1, enhanced the density, speed and magnetic field inside CME1. The peak magnetic field at 1 AU is about 69 nT, and the southward field component is as high as 50.5 nT. An intense geomagnetic storm with the minimum $D_{\rm st}$ index of $-221$ nT occurred despite a brief time period of the southward field. This can be considered as another manifestation of the ``perfect storm" scenario proposed by \cite{liu14a}, although the current case is not ``super" in the same sense as the 2012 July 23 event. 

\acknowledgments The research was supported by the Recruitment Program of Global Experts of China, NSFC under grants 41374173 and 41774179, and the Specialized Research Fund for State Key Laboratories of China. We acknowledge the use of data from SOHO, Wind, ACE, Ulysses and GOES.

\clearpage

\begin{figure}
\epsscale{0.9} \plotone{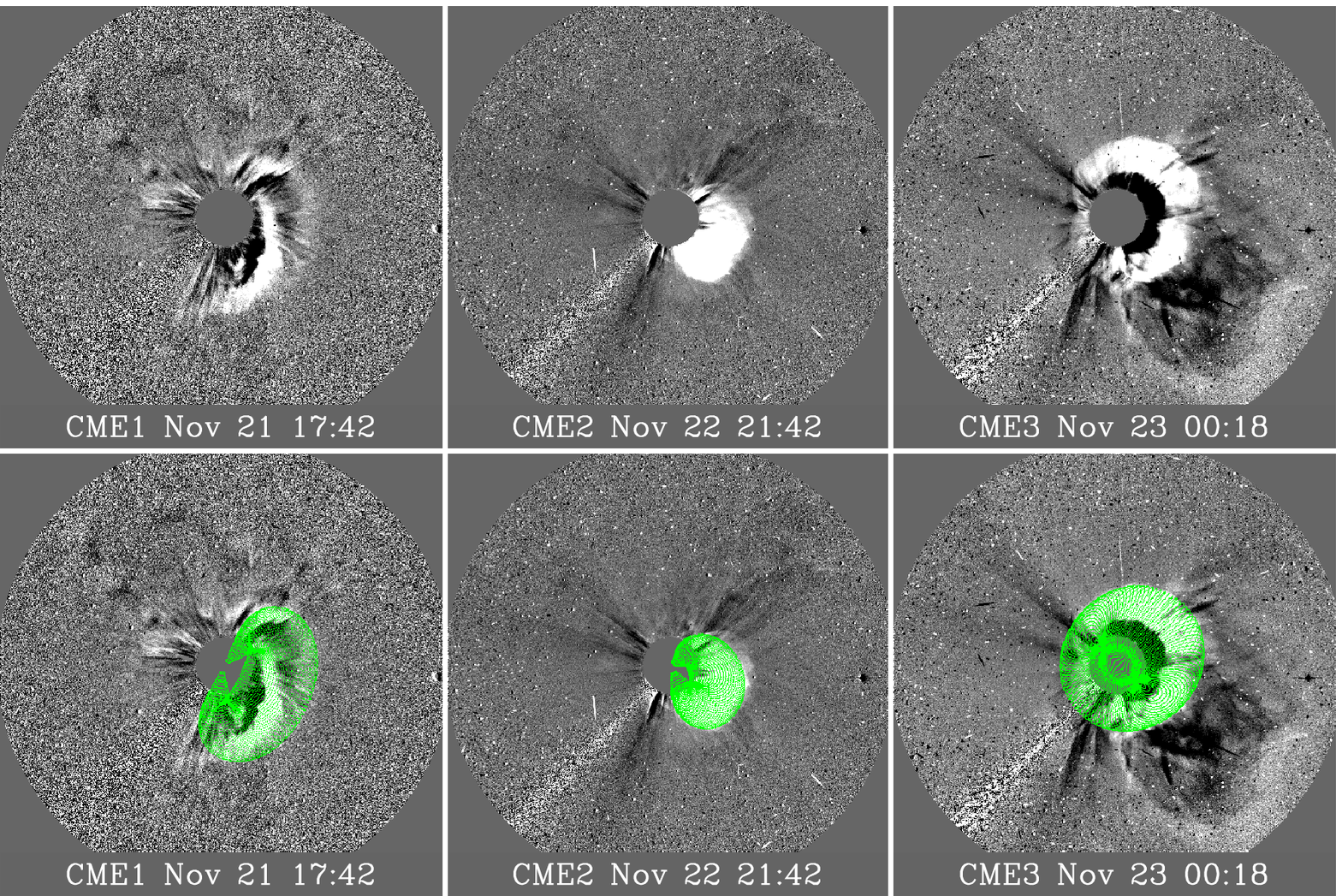} 
\caption{Top: running-difference images of the three CMEs from LASCO C3 of SOHO. Bottom: GCS wireframe rendering of the CMEs superposed on the observed images. Note that at 00:18 UT on November 23 both CME2 and CME3 were visible in LASCO C3.}
\end{figure}

\clearpage

\begin{figure}
\epsscale{0.8} \plotone{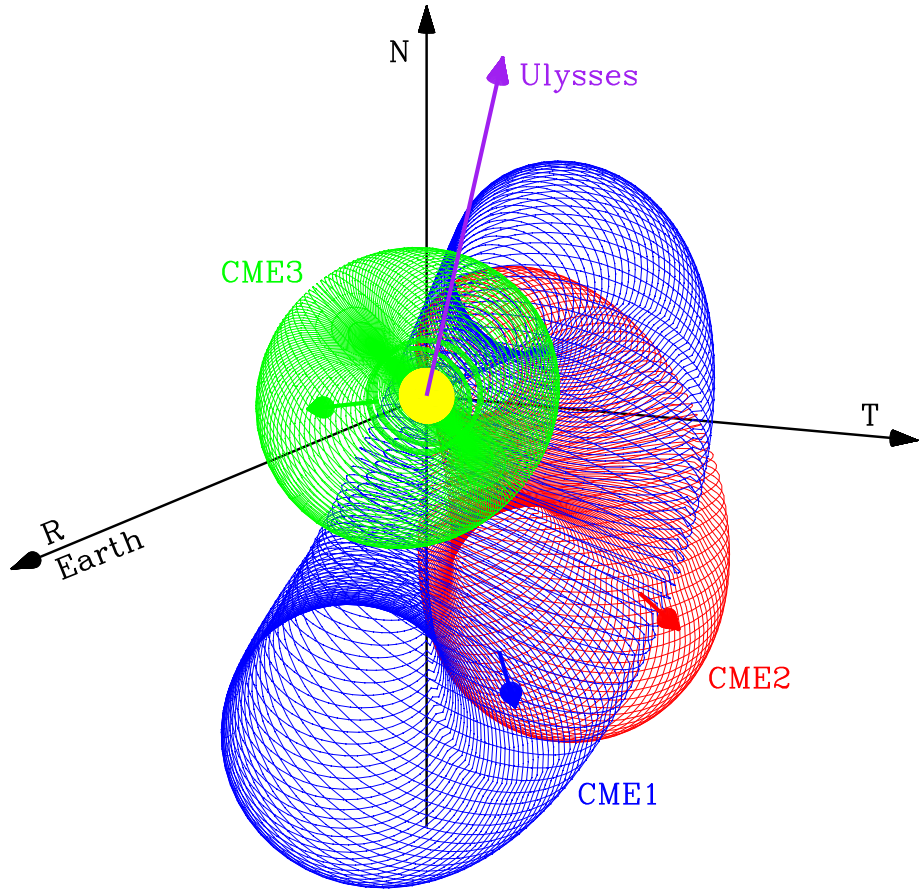} 
\caption{A 3D rendering of the three CMEs in RTN coordinates of the Earth. The yellow sphere is the Sun. The Earth is along the $\textbf{R}$ direction. The direction of Ulysses, which is about $57.9^{\circ}$ west and $71.4^{\circ}$ north of the Earth, is indicated by the purple arrow. The blue, red and green arrows mark the propagation directions of the three CMEs, respectively. The 3D structure is taken from the GCS modeling at 19:42 UT on November 21 for CME1, 22:18 UT on November 22 for CME2, and 23:42 UT on November 22 for CME3.}
\end{figure}

\clearpage

\begin{figure}
\epsscale{0.7} \plotone{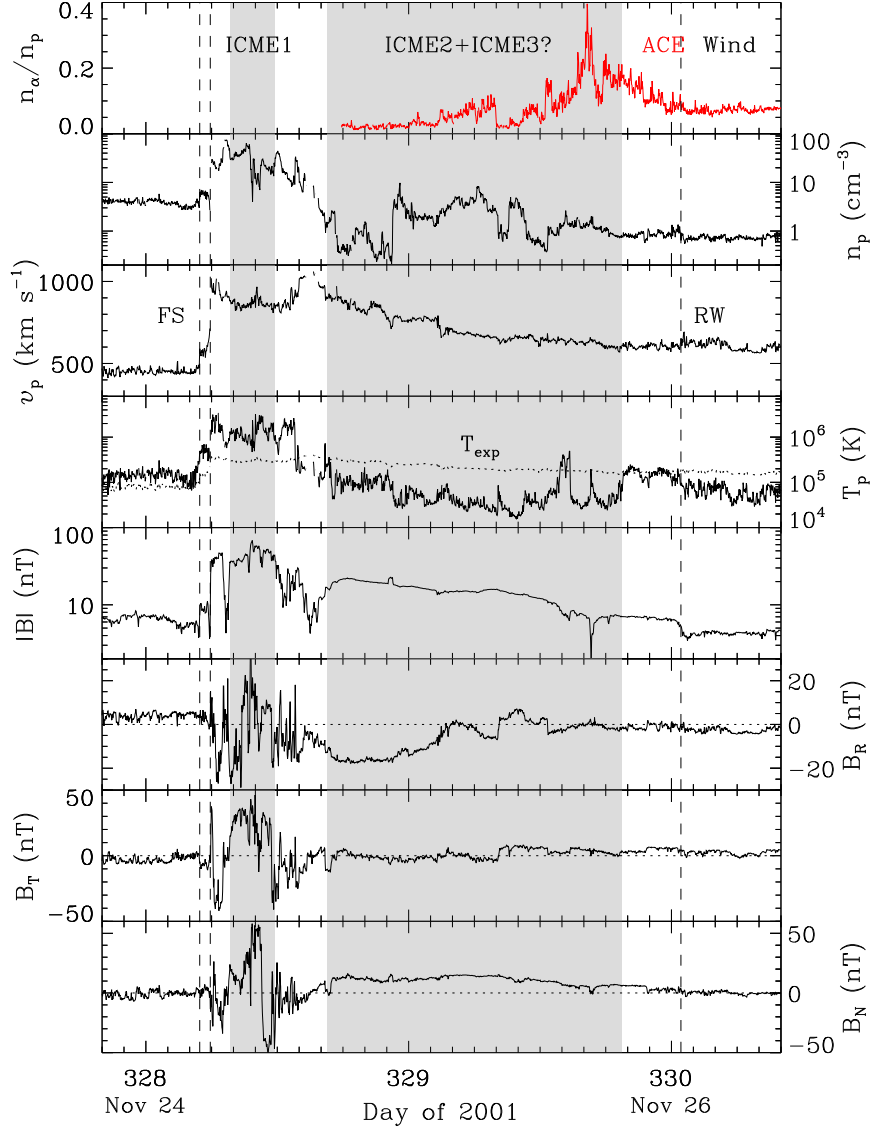} 
\caption{Solar wind measurements from Wind (black) and ACE (red) near the Earth. From top to bottom, the panels show the alpha-to-proton density ratio, proton density, bulk speed, proton temperature, and magnetic field strength and components, respectively. The shaded regions represent the ICME intervals. The first two vertical dashed lines mark two forward shocks (FS), and the third vertical dashed line indicates a reverse wave (RW). The dotted curve in the fourth panel denotes the expected proton temperature calculated from the observed speed \citep{lopez87}. As explained in the text, the first shock is driven by CME1, and the second one is originally from CME3 and has passed to the front of the system.}
\end{figure}

\clearpage

\begin{figure}
\epsscale{0.7} \plotone{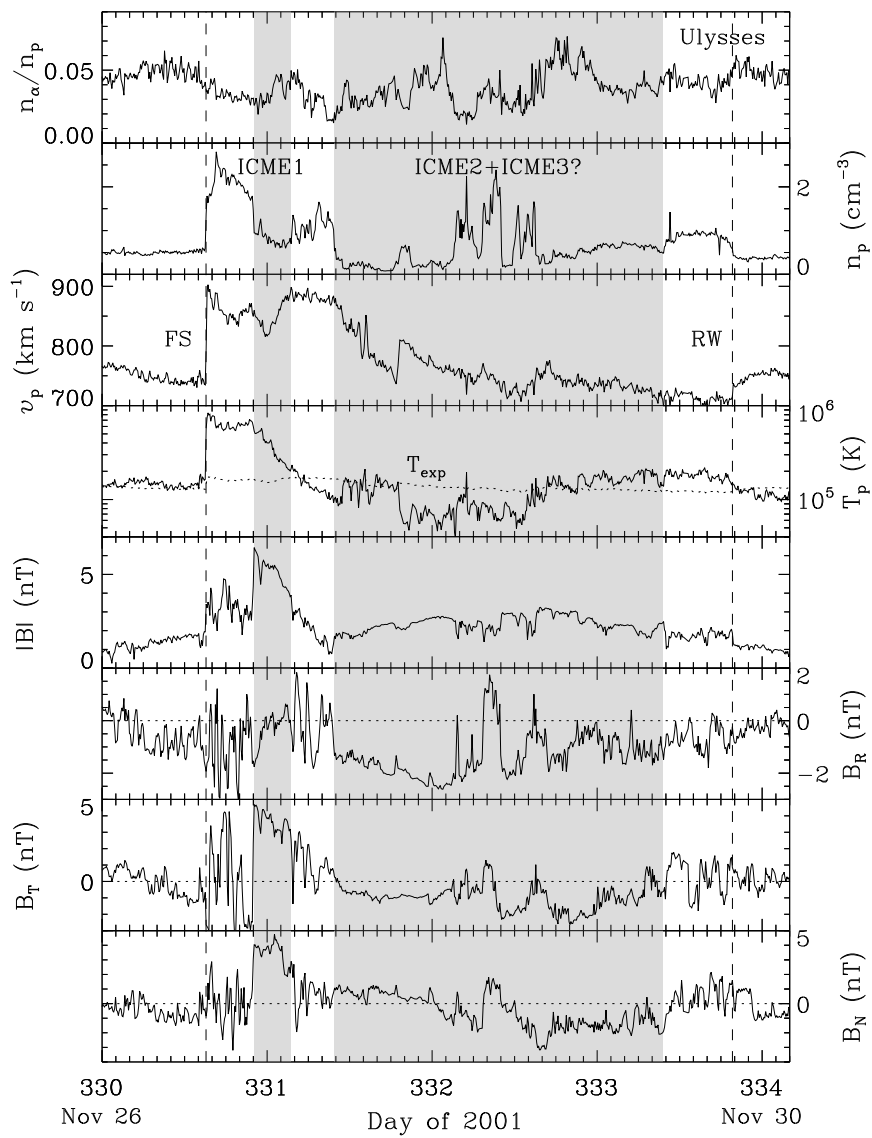} 
\caption{Similar to Figure~3, but for the solar wind measurements from Ulysses at 2.34 AU. The plasma and magnetic field profiles resemble those observed at the Earth. The calculation of the expected proton temperature takes into account a temperature-distance gradient of $r^{-0.7}$, where $r$ is the distance of Ulysses from the Sun.}
\end{figure}

\clearpage

\begin{figure}
\epsscale{0.7} \plotone{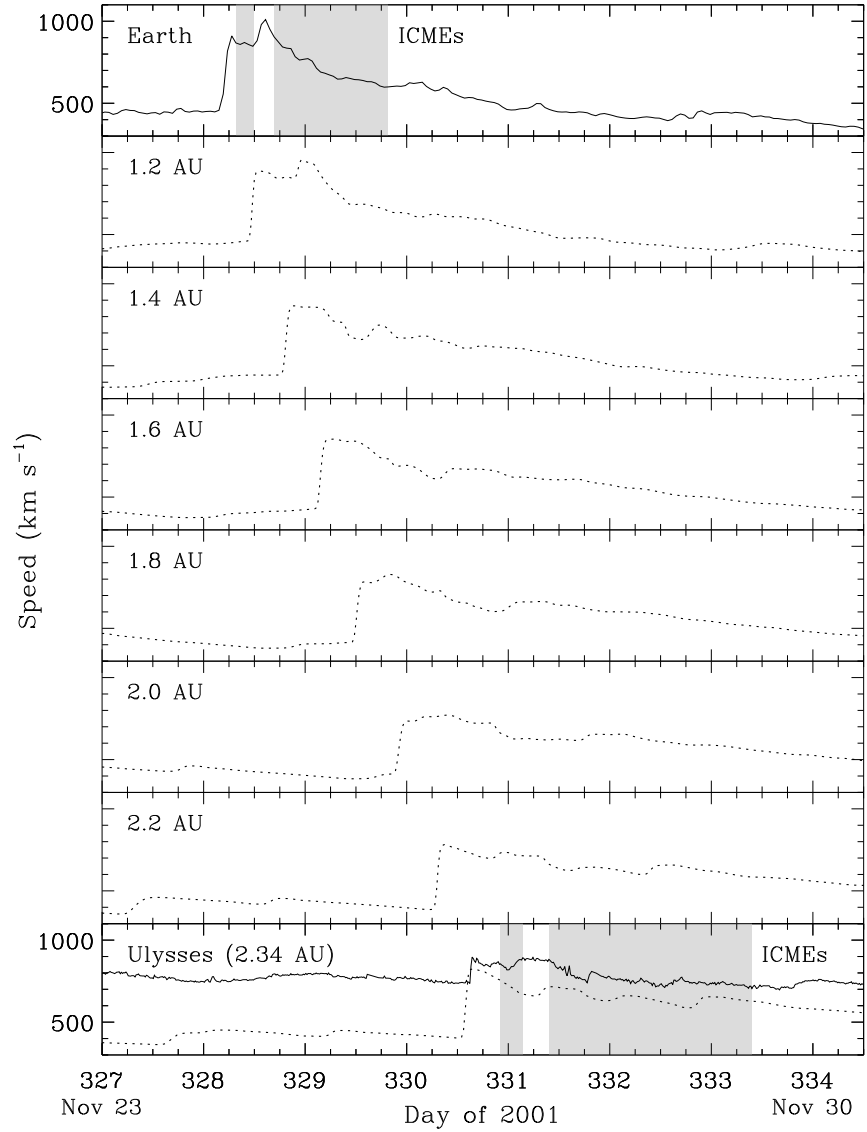} 
\caption{Evolution of solar wind speeds from the Earth to Ulysses via the MHD model. The solid curves in the top and bottom panels are the observed speeds at the Earth and Ulysses, respectively, and the dotted curves indicate the predicted speeds at given distances. The shaded regions represent the observed ICME intervals at the Earth and Ulysses. The same scale is used for the vertical axes of all the panels.}
\end{figure}

\clearpage

\begin{figure}
\epsscale{1.0} \plotone{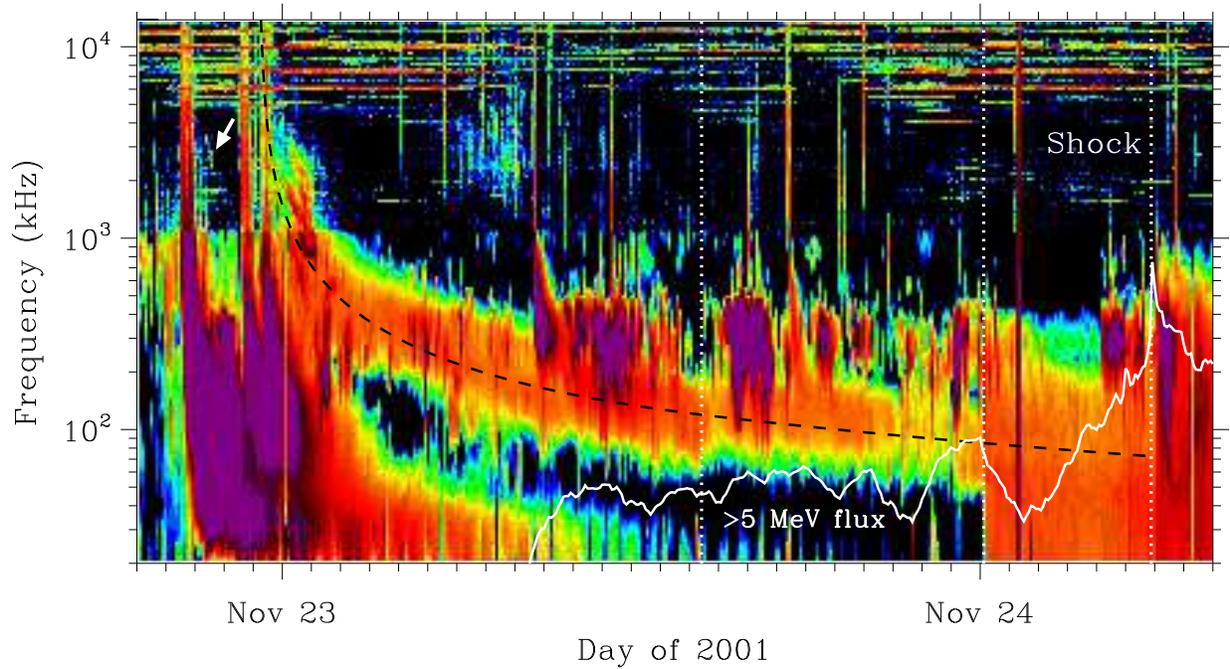} 
\caption{Dynamic spectrum from Wind. The black dashed curve, which is determined with a simple analytical model, simulates the frequency drift of the Sun-to-Earth type II burst. The first vertical dotted line marks the time (14:25 UT on November 23) when the shock slowed down to a roughly constant speed, the second vertical dotted line shows the time (00:07 UT on November 24) of a sudden broadening and enhancement in the type II burst, and the third vertical dotted line denotes the shock arrival time from in situ measurements at the Earth. The white solid curve represents the normalized intensity of $>$5 MeV particles observed by GOES. The white arrow indicates the type II burst associated with CME2.}
\end{figure}

\clearpage

\begin{figure}
\epsscale{0.8} \plotone{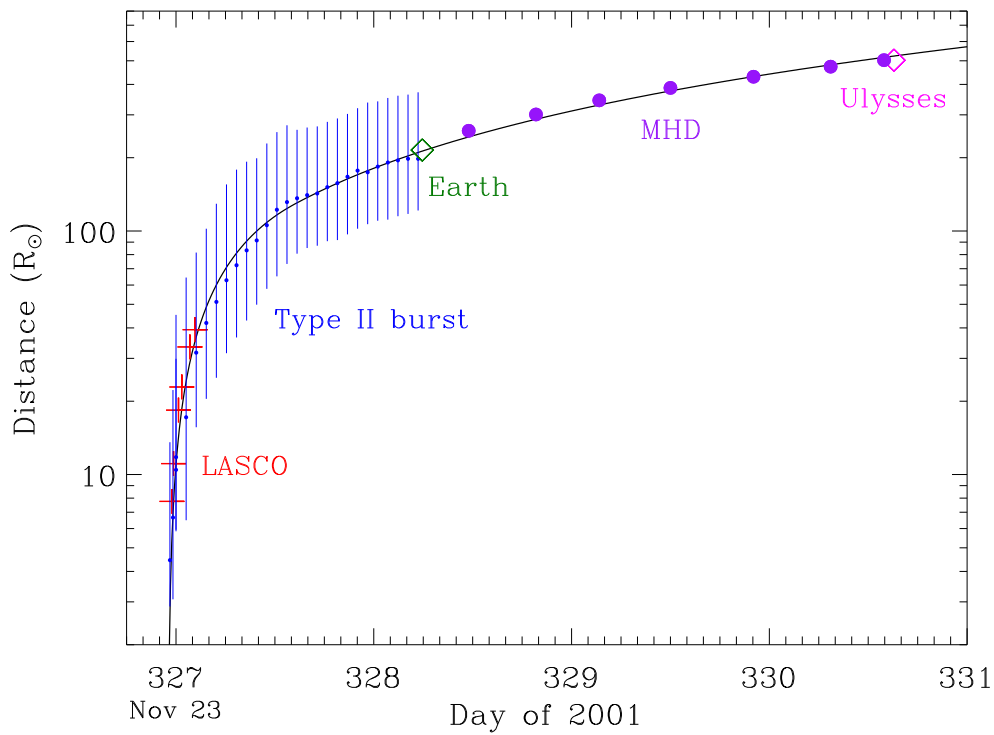} 
\caption{Overall shock propagation profile from the Sun far into interplanetary space determined from the analytical model (black solid curve). Crosses denote the LASCO observations near the Sun. Small dots show the distances obtained from the average frequencies of the Sun-to-Earth type II burst, with the error bars from the bandwidth of the type II burst. Diamonds indicate the shock arrival times at the Earth and Ulysses (2.34 AU). Between the Earth and Ulysses are the shock arrival times (filled circles) at 1.2, 1.4, 1.6, 1.8, 2.0, 2.2 and 2.34 AU given by the MHD model.}
\end{figure}

\end{document}